\documentclass[journal]{IEEEtran}

\usepackage{hyperref}
\usepackage{graphicx}
\usepackage{float}

\ifCLASSINFOpdf
\else

\fi
\usepackage{amsmath}
\usepackage{multirow}

\usepackage{algorithmic}
\usepackage{algorithm}
\usepackage{amssymb} 
\usepackage{float}      
\usepackage{needspace}   
\usepackage{capt-of}     
\usepackage{url}

\begin{document}
%
\title{LinkXplore: A Framework for Affordable High-Quality Blockchain Data}
%
%
%

\author{Peihao Li}

%
%

\markboth{LinkXplore framework preprint}%
{Shell \MakeLowercase{\textit{et al.}}: Bare Demo of IEEEtran.cls for IEEE Journals}
%



\maketitle

\begin{abstract}
Blockchain technologies are rapidly transforming both academia and industry. However, large-scale blockchain data collection remains prohibitively expensive, as many RPC providers only offer enhanced APIs with high pricing tiers that are unsuitable for budget-constrained research or industrial-scale applications, which has significantly slowed down academic studies and product development. Moreover, there is a clear lack of a systematic framework that allows flexible integration of new modules for analyzing on-chain data.

To address these challenges, we introduce LinkXplore, the first open framework for collecting and managing on-chain data. LinkXplore enables users to bypass costly blockchain data providers by directly analyzing raw data from RPC queries or streams, thereby offering high-quality blockchain data at a fraction of the cost. Through a simple API and backend processing logic, any type of chain data can be integrated into the framework. This makes it a practical alternative for both researchers and developers with limited budgets. Code and dataset used in this project are publicly available at \url{https://github.com/Linkis-Project/LinkXplore}
\end{abstract}


%
\IEEEpeerreviewmaketitle

\section{Introduction}
%
%
%
%
\IEEEPARstart{B}{lockchain} research has expanded significantly in recent years, with bibliometric surveys reporting growth in publication volume and a widening scope from consensus and cryptography to sector-specific applications such as supply chain, healthcare, and finance~\cite{LopezSorribes2023Sensors,Guo2021FGCS,Habil2024JCSS,Su2024PPNA}. Large-scale mappings report tens of thousands of publications since 2016, strong annual growth, broader international collaboration among authors, and a shift toward reusable datasets and comparative evaluations~\cite{LopezSorribes2023Sensors,Habil2024JCSS,Bao2025ArXiv}. Focused reviews trace how subfields such as industrial digital transformation and consensus design have matured~\cite{Su2024PPNA,MDPI2024InformationConsensus}. On the market side, milestones driven by regulation reshaped the crypto industry's landscape: the U.S. SEC approved eleven spot Bitcoin ETFs in 2024, catalyzing rapid growth in assets under management (AUM) and liquidity, with early empirical work documenting measurable effects on crypto returns and volatility~\cite{SEC2024SpotETF,CRS2024BitcoinETPs,Espel2024ETFSeasonality,Babalos2024SpotETFImpact}. Participation in public markets also broadened as firms that are native to the crypto industry entered public listings and large corporations allocated to digital assets (e.g., Circle’s listing and corporate treasury strategies by firms such as MicroStrategy), reinforcing institutional participation and investment. In the U.S., the GENIUS Act (2025) was signed into law, establishing a federal framework for payment stablecoins with issuer supervision, redemption rights, and consumer protection measures~\cite{Congress2025GENIUSActText,CRS2025GENIUSOverview,HFSC2025GENIUSOnePager,Reuters2025GENIUSSenatePass}. In the EU, the Markets in Crypto-assets (MiCA) framework began to apply to stablecoins in 2024, establishing standardized issuance and disclosure rules~\cite{EBA2024MiCAStablecoinStart,ESMA2025StablecoinsTransition}.

Despite this growth, both industry and academia remain constrained by data frictions. For data access, teams must either run large, often archive, nodes and keep them in sync or rely on external APIs with rate limits and fees. Raw ledgers are low level in general and they require ABI or IDL decoding, which raises the learning curve and makes analyses harder to reproduce~\cite{Mafrur2025BlockchainAnalytics,Wei2021SurveyBDMS,Kalodner2020BlockSci}. In terms of accuracy and validity, scripts written by users and imperfect pipelines often lead to double counting, missing events, and inconsistencies after protocol or contract upgrades. Address labels are unreliable, because different providers assign different tags, so metrics need cross validation~\cite{Mafrur2025BlockchainAnalytics,Azad2024MLforBlockchain}. Finally, even extracting ``just a few'' historical data points with quality guarantees is costly: Ethereum archive infrastructure demands multiple terabytes of storage (\(\approx 12\,\mathrm{TB}\) on Geth, \( \approx 2\text{--}3\,\mathrm{TB}\) on Erigon, and growing) and long initial syncs, while chains with high throughput often require specialized streaming plugins or history back ends, and these costs are prohibitive for many academic groups. For instance, Solana gRPC can relieve researchers of much of the work of decoding events across numerous IDLs, but access is often expensive even for small request volumes, which limits systematic study~\cite{Geth2025ArchiveMode,Polygon2025ErigonArchive,QuickNode2025YellowstoneGRPC}. A further cost constraint appears even in seemingly simple tasks such as retrieving the price at a given time \(t\). For centralized exchanges, this is a straightforward and reliable HTTP call. On Solana, teams typically depend on external APIs that package liquidity and events recorded on the chain (for example Birdeye and Pyth). These services offer small free tiers, but access beyond those limits requires paid credits or subscriptions, and costs rise quickly for higher request volumes, which constrains both research and product development~\cite{Birdeye2025Pricing,Pyth2025Hermes}.

These constraints motivate an affordable, high quality blockchain data framework. Such a framework should solve four challenges:
\begin{itemize}
  \item \textit{Affordability}: focus on algorithm design rather than paid throughput. Study how third party ``black box'' APIs produce good results, and then reimplement efficient methods that achieve similar accuracy at lower cost, ideally through algorithmic optimizations.
  \item \textit{Data quality}: start from actual user needs. Ship a small set of high demand data types and queries first, define them clearly, and keep the formats stable. Prefer canonical sources, record where each value came from, and avoid outputs that add maintenance cost without real usage.
  \item \textit{Extensibility}: make it straightforward to add support for new chains, new APIs, and revised data formats by using connectors and clear versioning, so that existing pipelines continue to work when components change.
  \item \textit{Openness and deployability}: offer an open source core with pluggable storage and compute back ends, support both local and cloud deployments, and scale across multiple machines, so teams can avoid dependence on a single vendor and adjust the system to their own constraints.

\end{itemize}

We present \textit{LinkXplore}, a framework designed to meet the four requirements outlined above in a single, practical system. For affordability, LinkXplore reconstructs widely used endpoints from first principles using canonical ledger data. Cost optimizations are achieved by choosing between RPC sources, avoiding expensive calls when possible, and caching historical data. The framework guarantees data quality by treating reputable RPC and market data services as references and validating reconstructed answers against them on sampled windows. Decoders are generated from versioned ABIs or IDLs and include checks around protocol upgrades. For extensibility, LinkXplore exposes a single, chain agnostic API and a separate module for each chain. Modules handle block transactions and metadata extraction, data decoding, and log analysis against versioned ABIs or IDLs. Pipelines are declared in configuration, and each module includes tests that catch changes as protocols evolve. Finally, openness and deployability come from an open source core and a stable plugin interface, which allow deployments that range from a single machine to a scaled cluster.

\begin{figure*}[htbp]
  \centering
  \includegraphics[width=\textwidth]{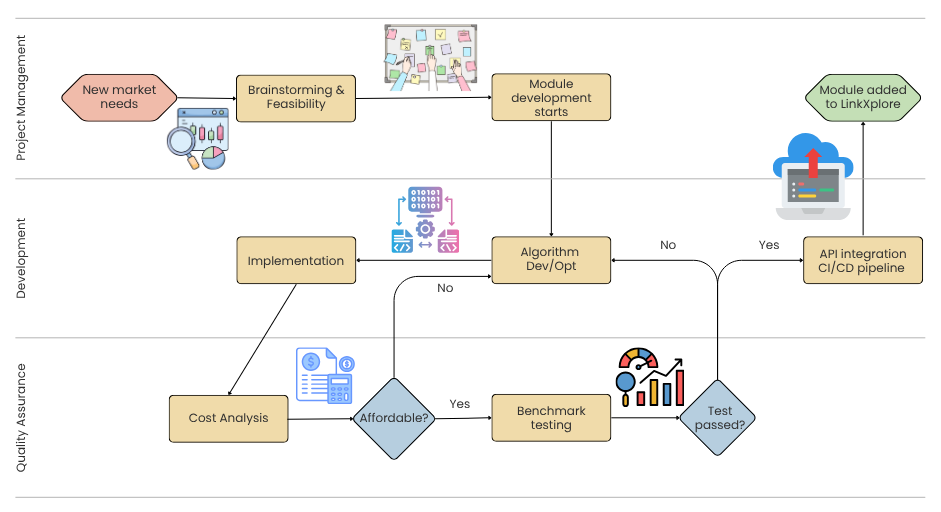}
  \caption{LinkXplore framework overview. The swimlanes (Project Management, Development, Quality Assurance, and Platform/DevOps) show how uniform, chain-agnostic APIs and per-chain adapters flow through cost planning, data-quality gates, and CI/CD to deliver deployable modules across chains.}
  \label{fig:linkxplore-framework}
\end{figure*}

\section{The Framework}

In general, extensibility, openness, and deployability can be achieved through appropriate software design patterns. However, \textbf{affordability} and \textbf{data quality} are the most challenging aspects in the absence of open solutions. Most such tools are kept private for profitability and are not open-sourced. They are typically used for private quantitative analysis or offered as paid third-party products, such as Birdeye~\cite{Birdeye2025Pricing}. If one lacks the budget for costly node access such as gRPC~\cite{QuickNode2025YellowstoneGRPC}, they must handle the dirty work of decoding binaries in transaction JSON streams. An open-source, well-structured framework with reasonable pricing would be of great interest to the blockchain/Web3 community.

Figure~\ref{fig:linkxplore-framework} summarizes how a market signal moves into a production module in \textit{LinkXplore}. The flow is organized into three lanes: project management, development, and quality assurance. The process validates the need, checks feasibility, manages costs, enforces a uniform, chain-agnostic API specification, and releases via CI/CD.

\begin{itemize}
  \item \textit{Project Management}: From \emph{New market needs} we derive a testable API (e.g., \url{/linkxplore/price/<chain>/<params>}) with target accuracy. In \emph{Brainstorming \& Feasibility}, the team confirms that the need is technically feasible and cost-optimizable, that performance can be reconstructed from canonical ledger data and the corresponding block explorer, and that test results provide sufficient validation. A go/no-go decision and a development plan lead to \emph{Module development begins}.

  \item \textit{Development}: In \emph{Algorithm Dev/Opt}, the team designs and develops the module’s algorithm, optimizing it subject to query-cost constraints. \emph{Implementation} builds the algorithm’s codebase. When the code passes cost and data-quality tests, the module advances to \emph{API integration + CI/CD}; otherwise, it iterates. The API query format remains identical across chains, so clients need only swap the chain segment in the query path.

  \item \textit{Quality Assurance}: \emph{Cost Analysis} models per-request costs across alternative execution plans. If the \emph{Affordable?} gate fails, the flow returns to \emph{Algorithm Dev/Opt}. If it passes, \emph{Benchmark testing} exercises representative workloads to measure latency and error rates. The \emph{Test passed?} decision requires accuracy against reputable references within reasonable tolerances; failures loop back to \emph{Algorithm Dev/Opt}.
\end{itemize}

Two properties sustain the flow: (i) efficient feedback loops—especially between cost analysis and algorithm design; and (ii) a stable public surface via the uniform API specification. Together, they enable extensibility through the adapter pattern and chain-agnostic endpoints, \textbf{affordability} through explicit cost gates and optimized execution plans, \textbf{data quality} through reference validation in benchmark testing gate.

To ensure \textbf{openness} and \textbf{deployability}, LinkXplore extends its adapter pattern with a plugin architecture and Kubernetes-based microservices. \textbf{Openness} comes from a standardized API contract (OpenAPI specifications) and a plugin interface (e.g., Go/Rust) for new endpoints, enabling third-party contributions without core code changes. Plugins handle chain-specific logic (e.g., \url{/linkxplore/<chain>/price}) with versioned ABIs/IDLs and rely on an open-source core (MIT license), comprehensive documentation, and CI/CD pipelines for plugin validation~\cite{OpenAPI2025Spec,Eclipse2023PluginModel}. \textbf{Deployability} is achieved by decomposing LinkXplore into microservices (a Core Service for routing, chain-specific services for Ethereum/Solana, and shared services for caching/logging), containerized with Docker and orchestrated via Kubernetes. Kubernetes resources (Pods, Services, Ingress, ConfigMaps/Secrets) support local (Minikube) and cloud (EKS/GKE) deployments, configured through Helm charts and YAML profiles for flexible setups~\cite{Kubernetes2025Docs,Helm2025Charts}. Observability uses Prometheus for metrics, ELK Stack for logs, and Jaeger for tracing, ensuring scalability and performance monitoring across environments while avoiding vendor lock-in~\cite{Prometheus2025Monitoring,Jaeger2025Tracing}.

\section{Adding a Market-Driven Module}\label{case_study}
This section walks through a complete iteration of the LinkXplore flow in Figure~\ref{fig:linkxplore-framework}, starting with a new market signal on Solana chain and ending with a deployed module that meets all four requirements. We begin in \emph{Project Management} by identifying a high demand data type with clear targets for accuracy and cost. We then move to \emph{Development}, where the algorithm is designed and implemented with a deterministic execution plan and an adapter that matches the uniform endpoint. In \emph{Quality Assurance}, cost models and reference checks validate affordability and data quality against reputable sources and test datasets. The module is then promoted to \emph{API integration + CI/CD}. The loop concludes with a released module that is affordable, accurate and deployable across environments. A successful implementation of such a module can prove that LinkXplore is useful and needed in the blockchain industry, and that it makes high quality data accessible for research and product development.

\subsection{Identifying market needs}
Developing products and doing research on the Solana chain is hard because mature tooling for decoding on-chain data is still limited \cite{Mafrur2025BlockchainAnalytics, QuickNode2025YellowstoneGRPC}. Even with good tools, unoptimized data access can become chaotic and expensive due to high RPC query costs. Unlike permissionless chains like Ethereum or Polygon, where anyone can sync the full chain for free \cite{Geth2025ArchiveMode,Polygon2025ErigonArchive}, participating in Solana validation requires staking large amounts of SOL to cover voting fees and operate sustainably \cite{solanaWhitepaper}. As a result, most applications and research on Solana rely on third-party RPC providers, including top products such as pump.fun and GMGN \cite{pumpfunDocs, gmgnDocs}. Premium providers such as QuickNode and Helius offer gRPC endpoints that free the users from low-level decoding works, but their pricing can discourage small teams and budget-constrained research projects \cite{QuickNode2025YellowstoneGRPC, HeliusPricingGRPC}. Without gRPC, users must decode Solana binary transaction data themselves by incorporating the different interfaces across programs.

We hereby propose a simple market need on the Solana chain: a pricing API. On centralized exchanges, it is easy to fetch large volumes of price data for a given token at time \(t\) through a simple API, but on decentralized exchanges it is difficult to obtain prices via an API without using third-party data providers. On Solana, for example, if a user wants the price of a token at time \(t\), they typically go to aggregator sites like Dexscreener or GMGN to view it in the UI, but getting it via HTTP requests is harder and often costs money. This reflects the nature of decentralized exchanges, where data that appears simple can be very expensive to access. Birdeye \cite{Birdeye2025Pricing} provides pricing via an API and lists token price as the first item in its API list, available in the free tier, which suggests that pricing best demonstrates core value at low cost and helps users activate within usage limits, indicating it is indeed a much-needed feature in the blockchain community \cite{KatoDumrongsiri2022ECRA, HsuYenHuNguyen2024ISEBM}. Moreover, one with a reliable price-fetching method can easily derive some of the paid API options of Birdeye, such as OHLCV-related APIs, which we will discuss later in the section.

\subsection{Algorithm development \& optimization}

\begin{enumerate}
\item \textbf{Cost optimization.} We do not use gRPC because that would force users to pay for premium access and make the framework too expensive to be useful. Instead we rely on standard RPC calls where you can send on the order of a few hundred requests per second without worrying much about cost. The tradeoff is that RPC returns raw transaction data, so we need good decoding logic to turn that raw data into clear information.

\item \textbf{Preliminary works.} The RPC node only returns encoded data for a queried transaction. We first need to build a reliable decoder that reads transactions and identifies swaps across common Solana AMMs and routers. For each trade, it answers two questions: which tokens are involved and how much was swapped. The decoder takes any transaction signature as input and returns swap information as output. Since swaps may occur on public or private programs, we ensure the decoder is compatible with most Solana AMMs and public routers and produces accurate results across different scenarios.

\item \textbf{Pricing at time $t$.} Map $t$ to the nearest slot, fetch that block, and filter to transactions that reference the target token using program ID and token account checks. Decode these to find swaps of the token for SOL, the most common base currency in Solana DEX swaps. For each transaction, compute the token price in SOL and aggregate prices in the block with a trade size weighted average. If no such trades are found in that slot, step back slot by slot until one appears or a set limit is reached. Return the result per HTTP query parameters, optionally with metadata.

\end{enumerate}

\subsection{Implementation}
\subsubsection{Swap Utility: Decoding Solana Swap Transactions}

This utility is a prerequisite for the pricing module. The pricing module queries historical prices by slot and needs a reliable decoder that turns raw transactions into a single \texttt{SwapInfo} per swap. Given a confirmed pair $(tx,\,meta)$, the utility prefers authoritative router or event evidence and falls back to leg aggregation only if such evidence is absent. Let $AK$ be the concatenation of message account keys and loaded addresses so that every index in instructions resolves to a public key. For clarity, an \emph{outer instruction} is an entry of \texttt{message.instructions} and an \emph{inner instruction} refers to \texttt{meta.innerInstructions} at the same outer index.

For each transaction we build two lookups
\[
TInfo:\ \text{acct}\mapsto(\text{mint},\mathrm{dec}), 
\qquad
Decs:\ \text{mint}\mapsto \mathrm{dec},
\]
primarily from \texttt{postTokenBalances} in $meta$ and seeded by Token and Token 2022 \texttt{Transfer} (opcode $3$) and \texttt{TransferChecked} (opcode $12$). We set $Decs[\text{SOL}]=9$ and default unseen mints to $0$. The effective signer is $s=AK[0]$, except for Jupiter DCA where $s=AK[2]$. A direction sanity step uses
\[
\Delta_{\text{SOL}}(s)=\text{postBalances}[s]-\text{preBalances}[s]
\]
to enforce that if $\Delta_{\text{SOL}}(s)>0$ and $\text{TokenIn}=\text{SOL}$, then input and output are swapped since the signer increased lamports.

Evidence is harvested only from the inner instruction set whose index matches an outer instruction $i$. The priority order is
\[
\begin{aligned}
\text{Jupiter Route event} &\ \succ\ \text{OKX log aggregate}\\
                          &\ \succ\ \text{Pump.fun event}\\
                          &\ \succ\ \text{leg aggregation}.
\end{aligned}
\]

Jupiter evidence is the Anchor RouteV2 event with a sixteen byte discriminator. OKX evidence uses router logs exposing \((\Delta_{\text{src}},\Delta_{\text{dst}})\) and reads mints from outer accounts. Pump.fun evidence is the Anchor trade event; if missing, use buy or sell discriminator and aggregate signer authorized \texttt{TransferChecked} at index. If no router evidence exists, scan Raydium, Orca, Meteora, and Pump.fun AMM outers, collect Token transfers, sum by mint, deduplicate via a stable leg key, take decimals from \(Decs\), and return \texttt{SwapInfo}.

\begin{algorithm}[H]
\caption{Swap Utility: Decoding a Solana Swap}
\begin{algorithmic}[1]\footnotesize
\REQUIRE transaction $tx$, metadata $meta$
\STATE $AK \gets$ message keys $\cup$ loaded addresses
\STATE Build $TInfo:\ \text{acct}\mapsto(\text{mint},\mathrm{dec})$ from $meta.\texttt{postTokenBalances}$; seed using Token and Token 2022 opcodes $3$ and $12$
\STATE Build $Decs:\ \text{mint}\mapsto \mathrm{dec}$ from $meta.\texttt{postTokenBalances}$; set unseen to $0$ and $Decs[\text{SOL}]\gets 9$
\STATE $swaps\gets[\ ]$;\ $found\gets\text{false}$
\FOR{each outer instruction $i$ in $tx$}
  \STATE $pid\gets AK[\text{ProgramIDIndex}(i)]$
  \IF{$pid=\texttt{JUPITER}$}
    \IF{some inner under $i$ starts with an eight byte RouteV2 event discriminator}
      \STATE Decode event; append tag \texttt{JUPITER}; $found\gets\text{true}$
    \ELSE
      \STATE Append AMM legs or Token transfers under $i$; if any then $found\gets\text{true}$
    \ENDIF
  \ELSIF{$pid=\texttt{OKX}$ router}
    \STATE $agg\gets$ parse router logs to obtain $(\Delta_{\text{src}},\Delta_{\text{dst}})$ and infer $(m_{in},m_{out})$ from outer accounts
    \IF{$agg\neq\varnothing$} \STATE Append \texttt{OKX} aggregate \ENDIF
    \STATE Append AMM legs under $i$; if any then $found\gets\text{true}$
  \ELSIF{$pid\in\{\texttt{Pump.fun program},\ \texttt{Pump.fun AMM}\}$}
    \IF{some inner under $i$ starts with an eight byte Pump.fun trade event discriminator}
      \STATE Decode; append \texttt{PUMP\_FUN}; $found\gets\text{true}$
    \ELSE
      \IF{outer data starts with the eight byte buy or sell discriminator}
        \STATE Aggregate signer authorized \texttt{TransferChecked} under $i$; if consistent then append and set $found\gets\text{true}$
      \ENDIF
    \ENDIF
  \ELSIF{$pid$ is a known bot router}
    \STATE Append legs by sweeping known AMM programs under index $i$
  \ENDIF
\ENDFOR

\IF{$found=\text{false}$}
  \FOR{each outer with $pid\in\{\texttt{Raydium},\texttt{Orca},\texttt{Meteora},\texttt{Pump.fun}\}$}
    \STATE Append Token transfers under that index
  \ENDFOR
\ENDIF

\STATE Partition $swaps$ into $JupEv,\ OkxAgg,\ PumpEv,\ Other$
\IF{$JupEv\neq\varnothing$}
  \STATE Aggregate by route level input and output mints to $(in,out,\mathrm{dec})$; \RETURN \texttt{SwapInfo}
\ENDIF
\IF{$OkxAgg\neq\varnothing$}
  \STATE Use $(in,out,\mathrm{dec})$ from the aggregate; \RETURN \texttt{SwapInfo}
\ENDIF
\IF{$PumpEv\neq\varnothing$}
  \STATE Map by $IsBuy$ with $\mathrm{dec}_{\text{SOL}}=9$; \RETURN \texttt{SwapInfo}
\ENDIF

\STATE Aggregate $Other$ by mint: for each mint $m$ sum unique legs $\sum \alpha_m$ with de duplication by leg key; prefer signer authorized outflow; otherwise choose the first and last unique mints as $(in,out)$
\STATE $s\gets AK[0]$; if the transaction contains Jupiter DCA then $s\gets AK[2]$
\STATE $\Delta_{\text{SOL}}(s)\gets meta.\texttt{postBalances}[s]-meta.\texttt{preBalances}[s]$
\IF{$\Delta_{\text{SOL}}(s)>0$ and $\text{TokenIn}=\text{SOL}$}
  \STATE Swap input and output
\ENDIF
\STATE \RETURN \texttt{SwapInfo} with mints, amounts, decimals from $Decs$, list of AMM tags, $s$, signatures, and a timestamp
\end{algorithmic}
\end{algorithm}

\texttt{SwapInfo} contains mints, raw integer amounts, per mint decimals, set of AMM names, signer, signatures, and a timestamp.

\subsubsection{Price Utility: Price at timestamp \texorpdfstring{$t$}{t}}

Given a target mint \(X\) and a UTC timestamp \(t\), return the price of \(X\) in a base \(B\in\mathcal{B}\) with
\[
\mathcal{B}=\{\mathsf{SOL},\mathsf{USDC},\mathsf{USDT}\},\qquad
\mathsf{SOL}\succ\mathsf{USDC}\succ\mathsf{USDT}.
\]
The utility maps \(t\) to the \emph{nearest} finalized Solana slot and then performs an \emph{efficient mint based prefilter} of that block’s transactions before decoding. If the chosen slot(s) contain no eligible swaps for \(X\), it backs off to earlier slots until a valid price is obtained or a cap is reached.

\paragraph*{Timestamp \(\to\) nearest slot.}
Let \(S_{\mathrm{now}}\) be the current finalized slot and \(t_{\mathrm{now}}=\mathrm{getBlockTime}(S_{\mathrm{now}})\).
Use a calibrated average block time \(\bar{\Delta}\) stored in a local index (default \(0.4\,\mathrm{s}\)) to form an initial guess
\[
\widehat{S}_0=S_{\mathrm{now}}-\mathrm{round}\!\Big(\frac{t_{\mathrm{now}}-t}{\bar{\Delta}}\Big).
\]
Find a bracket \([\!L,H\!]\) around \(\widehat{S}_0\) such that
\(\mathrm{getBlockTime}(L)\le t\le \mathrm{getBlockTime}(H)\).
Binary search this bracket to obtain
\[
\begin{aligned}
S_{\lfloor\cdot\rfloor} &= \max\{\,s:\ \mathrm{getBlockTime}(s)\le t\,\},\\
S_{\lceil\cdot\rceil}   &= \min\{\,s:\ \mathrm{getBlockTime}(s)\ge t\,\}.
\end{aligned}
\]
Pick the \emph{nearest} slot
\[
S^\star=\arg\min_{s\in\{S_{\lfloor\cdot\rfloor},\,S_{\lceil\cdot\rceil}\}} \big|\mathrm{getBlockTime}(s)-t\big|.
\]
If both are equidistant, treat \(\{S^\star\}\) as the \emph{pair} \(\{S_{\lfloor\cdot\rfloor},S_{\lceil\cdot\rceil}\}\) and combine their transactions.

\paragraph*{Per trade price and aggregation with base normalization.}
Let \(R_{C\to B}(t)\) denote the contemporaneous conversion rate at \(t\) between currencies \(C,B\in\mathcal{B}\), obtained from a deep liquidity Solana DEX pool (e.g., Raydium or Orca) near slot \(S^\star\); if unavailable, use a reputable oracle or centralized exchange midquote, and record the source. Each decoded swap \(i\) has
\[
\texttt{SwapInfo}_i=\bigl(m^{(i)}_{\mathrm{in}},a^{(i)}_{\mathrm{in}},\mathrm{dec}^{(i)}_{\mathrm{in}};\ 
m^{(i)}_{\mathrm{out}},a^{(i)}_{\mathrm{out}},\mathrm{dec}^{(i)}_{\mathrm{out}};\ \cdots\bigr).
\]
Normalize \(q(a,\mathrm{dec})=a/10^{\mathrm{dec}}\); set \(q^{(i)}_{\mathrm{in}}=q(a^{(i)}_{\mathrm{in}},\mathrm{dec}^{(i)}_{\mathrm{in}})\) and \(q^{(i)}_{\mathrm{out}}=q(a^{(i)}_{\mathrm{out}},\mathrm{dec}^{(i)}_{\mathrm{out}})\), and treat wSOL as SOL. If swap \(i\) is quoted versus \(C\in\mathcal{B}\),
\[
\begin{aligned}
p_i^{(C)} &=
\begin{cases}
\dfrac{q^{(i)}_{\mathrm{out}}}{q^{(i)}_{\mathrm{in}}}, & (m^{(i)}_{\mathrm{in}},m^{(i)}_{\mathrm{out}})=(X,C),\\[0.6ex]
\dfrac{q^{(i)}_{\mathrm{in}}}{q^{(i)}_{\mathrm{out}}}, & (m^{(i)}_{\mathrm{in}},m^{(i)}_{\mathrm{out}})=(C,X),\\
\text{undef.}, & \text{otherwise.}
\end{cases}
\\[1ex]
w_i^{(C)} &=
\begin{cases}
q^{(i)}_{\mathrm{out}}, & m^{(i)}_{\mathrm{out}}=C,\\
q^{(i)}_{\mathrm{in}}, & m^{(i)}_{\mathrm{in}}=C.
\end{cases}
\end{aligned}
\]

Convert each trade to the requested base \(B\) using the rate matrix:
\[
p_i^{(B)} = p_i^{(C)} \cdot R_{C\to B}(t),\qquad
w_i^{(B)} = w_i^{(C)} \cdot R_{C\to B}(t).
\]
Apply base specific dust \(w_i^{(B)}\ge\tau_B\) (e.g., \(\tau_{\mathsf{SOL}}=10^{-4}\), \(\tau_{\mathsf{USDC}}=10^{-2}\)). Let \(\tilde p^{(B)}=\mathrm{median}\{p_i^{(B)}\}\) over kept trades; keep trades with \(\bigl|\log p_i^{(B)}-\log \tilde p^{(B)}\bigr|\le\log r\) for \(r\in[1.5,2]\). The slot or two slot volume weighted average price is
\[
P^{(B)}=\frac{\sum_i w_i^{(B)}\,p_i^{(B)}}{\sum_i w_i^{(B)}}.
\]

\begin{algorithm}[H]
\caption{Price Utility: Price at timestamp t via nearest slot search, base normalization, and filtered decoding}
\begin{algorithmic}[1]\footnotesize
\REQUIRE target mint X, UTC timestamp t; bases \(\mathcal{B}=[\mathsf{SOL},\mathsf{USDC},\mathsf{USDT}]\); thresholds \(\tau_B\); fence \(r\); max backoff K; calibrated \(\bar{\Delta}\) (default 0.4s)
\STATE \(S_{\mathrm{now}}\gets \texttt{getSlot}(\text{finalized})\); \(t_{\mathrm{now}}\gets \texttt{getBlockTime}(S_{\mathrm{now}})\)
\STATE \(\widehat{S}_0\gets S_{\mathrm{now}}-\mathrm{round}\!\big((t_{\mathrm{now}}-t)/\bar{\Delta}\big)\)
\STATE Find \([L,H]\) around \(\widehat{S}_0\) with \(\texttt{getBlockTime}(L)\le t\le \texttt{getBlockTime}(H)\)
\STATE Binary search \([L,H]\) to get \(S_{\lfloor\cdot\rfloor}\) and \(S_{\lceil\cdot\rceil}\); set \(C\) to the nearest slot(s)
\FOR{for k = 0 to K}
  \STATE \(C_k\gets \{\,s-k:\ s\in C\,\}\)
  \STATE \(Txs\gets[\,]\)
  \FOR{each s in \(C_k\)}
     \STATE \(blk\gets \texttt{getBlock}(s,\ \text{full transactions + metadata})\)
     \FOR{each transaction u in blk}
        \STATE mint hit \(\gets\) any entry in u.\texttt{preTokenBalances} or \texttt{postTokenBalances} with \texttt{mint} \(=X\)
        \STATE program hit \(\gets\) outer \texttt{programId} in \{Raydium, Orca, Meteora, Jupiter, OKX router, Pump.fun AMM\}
        \IF{mint hit or program hit} \STATE append u to \(Txs\) \ENDIF
     \ENDFOR
  \ENDFOR
  \STATE \(\mathcal{T}\gets[\,]\)
  \FOR{each tx in \(Txs\)} \STATE append \(\texttt{SwapInfo}(tx,meta)\) to \(\mathcal{T}\) if \(X\) is in \(\{m_{\mathrm{in}},m_{\mathrm{out}}\}\) \ENDFOR
  \STATE Build rate matrix \(R_{C\to B}(t)\) for \(C,B\in\mathcal{B}\) from a deep DEX pool near \(C_k\); if missing, use oracle or centralized midquote
  \FOR{each requested base \(B\) in \(\mathcal{B}\)}
     \STATE \(P\gets[\,]\); \(W\gets[\,]\)
     \FOR{each info in \(\mathcal{T}\)}
        \STATE parse \((m_{\mathrm{in}},a_{\mathrm{in}},\mathrm{dec}_{\mathrm{in}};\ m_{\mathrm{out}},a_{\mathrm{out}},\mathrm{dec}_{\mathrm{out}})\); coalesce wSOL to SOL
        \STATE determine pair currency \(C\in\mathcal{B}\) if \(\{m_{\mathrm{in}},m_{\mathrm{out}}\}=\{X,C\}\); else continue
        \STATE compute \(p^{(C)}\) and \(w^{(C)}\) as defined; convert \(p^{(B)}\gets p^{(C)} R_{C\to B}(t)\), \(w^{(B)}\gets w^{(C)} R_{C\to B}(t)\)
        \IF{\(w^{(B)}<\tau_B\)} \STATE continue \ENDIF
        \STATE append \(p^{(B)}\) to \(P\); append \(w^{(B)}\) to \(W\)
     \ENDFOR
     \IF{\(\sum W=0\)} \STATE continue \ENDIF
     \STATE \(\tilde p\gets \mathrm{median}(P)\); keep \(i\) with \(|\log P[i]-\log \tilde p|\le \log r\)
     \STATE \(vwap\gets \big(\sum_i W[i]\cdot P[i]\big)\big/\big(\sum_i W[i]\big)\)
     \STATE \textbf{return} \texttt{PriceInfo}\((vwap,\ B,\ \min C_k,\ \text{``slot\_vwap''},\ \text{len}(P),\ \sum W)\)
  \ENDFOR
\ENDFOR
\STATE widen to \([t-3600\text{ s},\,t]\) and recompute; if still unavailable, report \texttt{not\_available}
\end{algorithmic}
\end{algorithm}

If no base yields trades, back off to earlier slots and repeat. Base normalization ensures that swaps quoted in \(\mathsf{SOL}\), \(\mathsf{USDC}\), or \(\mathsf{USDT}\) contribute consistently to VWAP in the requested base via \(R_{C\to B}(t)\). Optional fallbacks include triangulation through the deepest available edge and short window aggregation.

So far we have successfully developed the pricing API with clear algorithmic implementations using Golang. To evaluate its performance, we will now proceed to the testing phase before integrating the LinkXplore framework.

\subsection{Cost analysis \& Benchmark testing}
\subsubsection{Evaluation Setup and Dataset}
For this newly developed pricing API module, we will use a common Solana RPC node provider, Chainstack, given its reputation for reliable service. All API queries will be made to this RPC provider, and the results will be transformed into pricing information using our developed algorithm, then compared against  the ground truth. API costs will be evaluated by summing the Chainstack gauge usage and the Birdeye usage dashboard so we can assess whether the total cost is acceptable compared to existing industry solutions.

We will compare the query cost using Birdeye, which is an excellent and widely used platform on Solana, especially for DEX-traded SPLs and newly launched tokens. In the evaluation section, we will use it as the ground truth and as a cost analysis comparison benchmark, since it is widely adopted by many small and medium sized businesses and products (e.g., SolanaUI, community projects, and Phantom, which has integrated Birdeye data). 

\subsubsection{Cost analysis}
To ensure integration with LinkXplore, we begin with a cost analysis of products currently in use and examine several major data-usage scenarios.

\begin{itemize}
\item \textbf{Real-Time Stream Processing} evaluates algorithms on real-time data, integrating pricing modules to analyze each relevant transaction.

Our module analyzes raw transactions and extracts prices directly, so analyzing ongoing transaction streams incurs only RPC-node costs. With Chainstack’s Solana RPC, block streaming is free and unlimited; thus, the module’s cost in this scenario is zero.

Birdeye is usage-based, so one pays for each transaction in the real-time stream. Users stream transactions from RPC nodes and then fetch relevant data from a pricing API. 
Let blocks arrive at rate $\lambda$ (blocks/s). Over $t$ hours the number of blocks is
\[
N \;=\; \lambda \cdot 3600 \cdot t.
\]
Let the analyzed-per-block count be $k \sim \mathcal{N}(\mu,\sigma^2)$ i.i.d. across blocks, so the total analyzed transactions are
\[
T \;=\; \sum_{i=1}^{N} k_i \;\sim\; \mathcal{N}\!\big(N\mu,\;N\sigma^2\big).
\]
Assume each analyzed transaction triggers $r$ pricing requests, each request consumes $c_{\mathrm{req}}$ compute units (CUs), and the CU price is $p_{\mathrm{CU}}$ USD/CU. The cost over $t$ hours is
\begin{equation}\label{eq:birdeye-cost}
C(t) \;=\; p_{\mathrm{CU}}\,c_{\mathrm{req}}\,r\,T.
\end{equation}
Hence
\begin{equation}\label{eq:birdeye-moments}
\begin{aligned}
\mathbb{E}[C(t)] \;&=\; p_{\mathrm{CU}}\,c_{\mathrm{req}}\,r\,\mathbb{E}[T]
\;=\; p_{\mathrm{CU}}\,c_{\mathrm{req}}\,r\,\mu\,\lambda\,3600\,t,\\
\mathrm{Var}\!\big(C(t)\big) \;&=\; \big(p_{\mathrm{CU}}\,c_{\mathrm{req}}\,r\big)^2\,\mathrm{Var}(T)\\
&=\; \big(p_{\mathrm{CU}}\,c_{\mathrm{req}}\,r\big)^2\,\sigma^2\,\lambda\,3600\,t,\\
\mathrm{SD}\!\big(C(t)\big) \;&=\; p_{\mathrm{CU}}\,c_{\mathrm{req}}\,r\,\sigma\,\sqrt{\lambda\,3600\,t}.
\end{aligned}
\end{equation}

A normal approximation gives the $95\%$ interval
\begin{equation}\label{eq:birdeye-ci}
C(t) \;\approx\; \mathbb{E}[C(t)] \;\pm\; 1.96\,\mathrm{SD}\!\big(C(t)\big).
\end{equation}

With $p_{\mathrm{CU}}=\tfrac{2}{60{,}000}$ USD/CU \cite{Birdeye2025Pricing}, $c_{\mathrm{req}}=64$, $r=1$, $\mu=10$, $\sigma=5$, and $\lambda=2.5$,
\[
\mathbb{E}[C(t)] \;=\; \frac{2}{60{,}000}\cdot 64 \cdot 1 \cdot 10 \cdot 2.5 \cdot 3600\,t
\;=\; 192\,t\ \text{USD},
\]
\[
\mathrm{SD}\!\big(C(t)\big) \;=\; \frac{2}{60{,}000}\cdot 64 \cdot 1 \cdot 5 \cdot \sqrt{2.5 \cdot 3600\,t}
\;\approx\; 1.01\,\sqrt{t}\ \text{USD}.
\]
For $t=1$ hour, $N=2.5\times 3600=9000$.
\[
\mathbb{E}[C(1)] = 192~\text{USD},\quad \mathrm{SD}\!\big(C(1)\big) \approx 1.01~\text{USD}.
\]
and a $95\%$ interval is $C(1)\in[190.0,\,194.0]$ USD.

In real-time stream processing, our pricing module costs \$0 regardless of duration, versus approximately \$190 per hour for current solutions, effectively eliminating costs.

\begin{figure}[H]
\centering
\includegraphics[width=\linewidth]{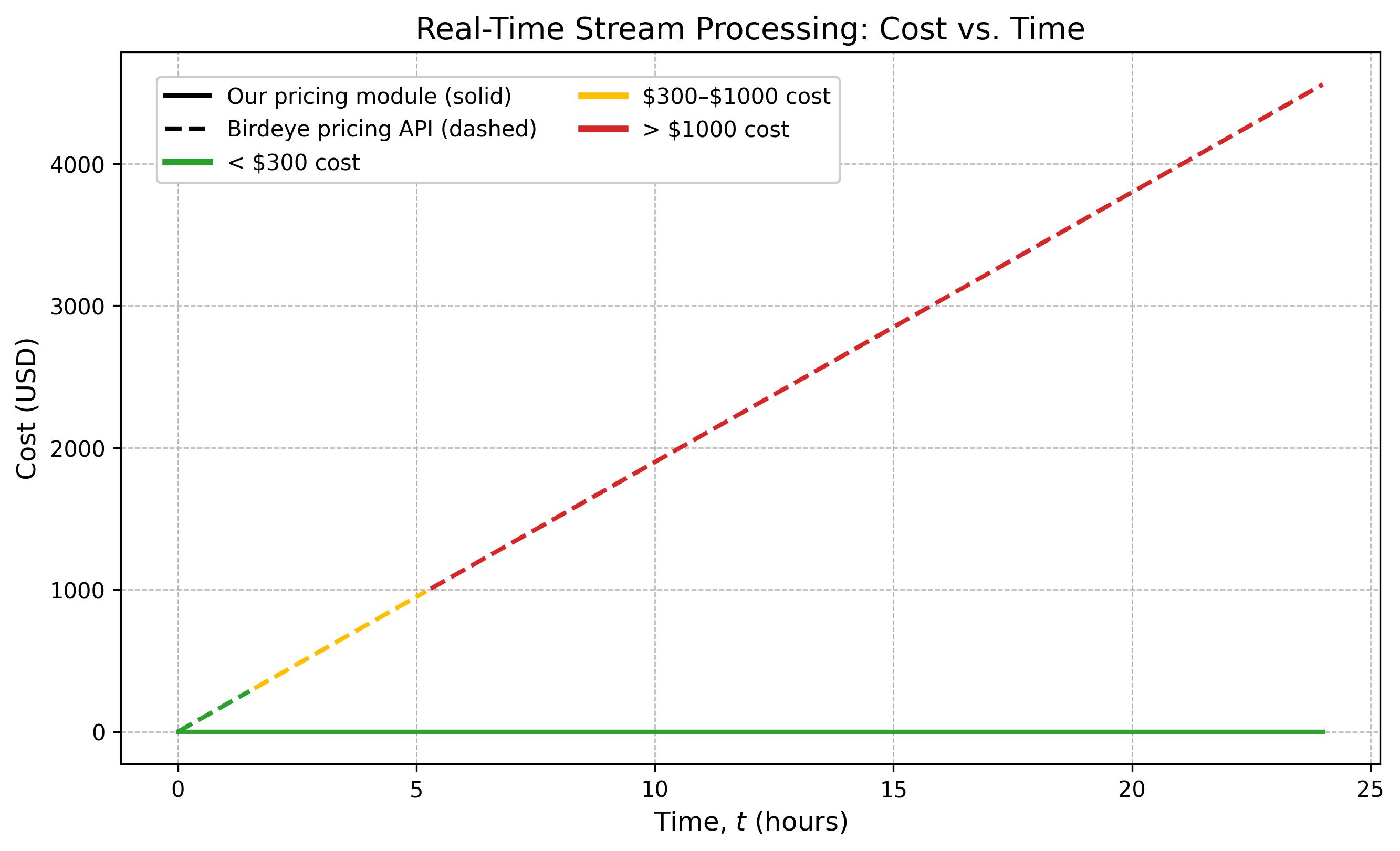}%
\caption{Real-time stream processing: our module costs \$0; Birdeye $\approx 190\,t$ USD. 
Line styles: solid = our module, dashed = Birdeye. Colors indicate cost bands: $<\$300$ (green), \$300–\$1000 (yellow), $>\$1000$ (red).}
\label{fig:cost_rt}
\end{figure}

\item \textbf{Major Event Analytics} evaluates recent high-impact events that usually last 1–2 days, with a 4–8 hour peak period when most blocks contain dense information across transactions. One representative case is the \$TRUMP token generation event (TGE), which triggered massive wealth creation \cite{TrumpToken} and drove strong demand for information extraction, for example analyzing price swings, social media sentiment, and identifying smart money addresses, where a price API is crucial for computing per address profit and loss (P\&L).

In our pricing module, for intensive analysis over a short period, we download all blocks from that interval rather than querying per transaction to avoid repeated fetches; thus, the cost is incurred by \texttt{getBlock} RPC method. Let $\rho_{\text{getBlock}}$ be the request units (RU) consumed per \texttt{getBlock} call. With block rate $\lambda$ (blocks/s) over $t$ hours (one block per slot), the RU consumption and optional dollar cost are
\begin{equation}\label{eq:getblock-ru-cost}
\begin{aligned}
\mathrm{RU}(t) &= \rho_{\text{getBlock}}\,\lambda\,3600\,t,\\
C_{\text{getBlock}}(t) &= p_{\text{RU}}\,\rho_{\text{getBlock}}\,\lambda\,3600\,t,
\end{aligned}
\end{equation}
where $p_{\text{RU}}$ denotes the USD per RU. In our case, $\rho_{\text{getBlock}}=1$ for blocks within 1.5 days and $\rho_{\text{getBlock}}=2$ otherwise. Under our Chainstack plan, access to up to two days of block (in the \$TRUMP TGE event) data is priced at \$1.35.

Birdeye must access every block to price a token, so the cost depends on the lookback period required to retrace the major event. Let $p_{\mathrm{CU}}$ be USD per compute unit (CU), $c_{\mathrm{req}}$ the CUs per Birdeye pricing request, $\lambda$ the block rate (blocks/s), over $t$ hours,
\begin{equation}\label{eq:birdeye-direct-fixed}
C_{\text{Birdeye}}(t) \;=\; p_{\mathrm{CU}}\, c_{\mathrm{req}} \cdot \lambda \cdot 3600 \cdot t.
\end{equation}
With $p_{\mathrm{CU}}=\tfrac{2}{60{,}000}$, $c_{\mathrm{req}}=64$, $\lambda=2.5$:
 $C_{\text{Birdeye}}(t)=19.2\,t~\text{USD}$.
 
Cost in 2 days ($t=48$):
\[
C_{\text{Birdeye}}(48) = 19.2 \times 48 = 921.6~\text{USD}.
\]

Results suggest that, in major event analysis, our pricing module consumes \$1.35 versus \$921.6 for current industry solutions, a 99.85\% cost reduction.

\begin{figure}[H]
\centering
\includegraphics[width=\linewidth]{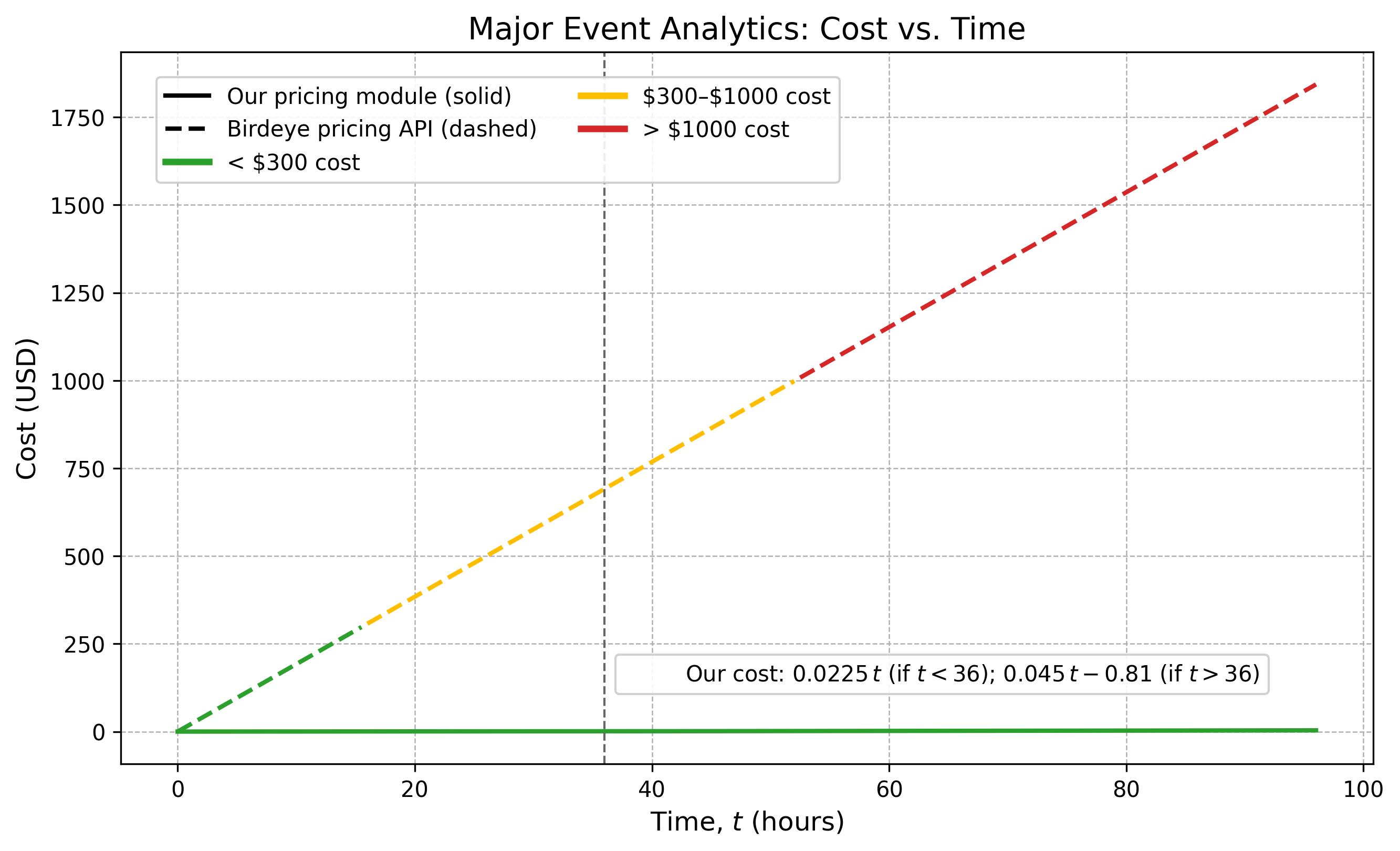}%
\caption{Major event analytics: our module $0.0225\,t$ (if $t<36$) and $0.045\,t-0.81$ (if $t>36$); Birdeye $19.2\,t$ USD. 
Line styles: solid = our module, dashed = Birdeye. Colors indicate cost bands: $<\$300$ (green), \$300–\$1000 (yellow), $>\$1000$ (red).}
\label{fig:cost_event}
\end{figure}
\end{itemize}

This section quantified the cost of price computation under two representative workloads and contrasted our module against a usage–metered baseline. We consider two major use cases of pricing APIs on the Solana chain, both with significant demand from the academic and industry sectors. For \emph{real-time stream processing}, our approach derives prices directly from raw transactions and leverages free block streaming on Chainstack, yielding zero marginal cost for any horizon; by contrast, the Birdeye-based pipeline grows linearly with time as \(C(t)\) in~\eqref{eq:birdeye-cost} with moments in~\eqref{eq:birdeye-moments} and interval~\eqref{eq:birdeye-ci}, which, under the standard parameterization (in a lightweight workload), implies \(\mathbb{E}[C(t)]\approx 190\,t\) USD (e.g., \(\mathbb{E}[C(1)]\approx\$192\)), as shown in Fig.~\ref{fig:cost_rt}. For \emph{major event analytics}, our design enables access by fetching entire blocks via \texttt{getBlock}, incurring RU–based spend \(C_{\text{getBlock}}(t)=p_{\text{RU}}\rho_{\text{getBlock}}\lambda\,3600\,t\) in~\eqref{eq:getblock-ru-cost}; within Chainstack’s two–day retention window this totals \(\$1.35\) for a complete backfill of the event interval, whereas a Birdeye-only workflow scales as \(C_{\text{Birdeye}}(t)=p_{\mathrm{CU}}c_{\mathrm{req}}\lambda\,3600\,t\) in~\eqref{eq:birdeye-direct-fixed} (e.g., \(\$921.6\) for \(t=48\) hours); see Fig.~\ref{fig:cost_event}. Results indicate that our module eliminates real-time streaming costs and reduces event-history analysis costs by roughly \(99.8\%\)–\(100\%\).

\subsubsection{Benchmark testing}
Since the pricing module is confirmed affordable, we evaluate accuracy against the Birdeye ground truth. For each asset \(X\) and each sample time \(t\in S_X\), let \(\widehat{P}_{X,t}\) denote the price returned by our module (slot-based VWAP as defined above) and let \(G_{X,t}\) denote the corresponding Birdeye price. Define the pointwise error \(e_{X,t}=\widehat{P}_{X,t}-G_{X,t}\). We report mean squared error (MSE) and the standard deviation (SD) of these errors both per asset and pooled across assets to summarize performance. In addition, we summarize percentage errors using the relative error \(r_{X,t}=e_{X,t}/G_{X,t}\) whenever \(G_{X,t}>0\).

\paragraph*{Per-asset metrics}
For a fixed asset \(X\) with \(n_X=|S_X|\),
\[
\mathrm{MSE}_X \;=\; \frac{1}{n_X}\sum_{t\in S_X}\!\bigl(e_{X,t}\bigr)^2,
\qquad
\bar e_X \;=\; \frac{1}{n_X}\sum_{t\in S_X}\! e_{X,t}.
\]
The sample standard deviation is
\[
s_X \;=\; \sqrt{\frac{1}{n_X-1}\sum_{t\in S_X}\! \Bigl(e_{X,t}-\bar e_X\Bigr)^2 }.
\]
For percentage errors we use mean absolute percentage error (MAPE), reported in percent. Let \(S_X^{(\%)}=\{\,t\in S_X:\,G_{X,t}>0\,\}\) and \(n_X^{(\%)}=|S_X^{(\%)}|\). Then
\[
\mathrm{MAPE}_X \;=\; 100\cdot\frac{1}{n_X^{(\%)}}\sum_{t\in S_X^{(\%)}}\! \lvert r_{X,t}\rvert.
\]

\paragraph*{Pooled metrics across assets}
Let \(\mathcal{X}\) be the set of evaluated assets and \(N=\sum_{X\in\mathcal{X}} n_X\). The pooled error metrics are
\begin{align*}
\mathrm{MSE}_{\mathrm{all}}
  &= \frac{1}{N}\sum_{X\in\mathcal{X}} \ \sum_{t\in S_X}\!\bigl(e_{X,t}\bigr)^2,\\
\bar e_{\mathrm{all}}
  &= \frac{1}{N}\sum_{X\in\mathcal{X}} \ \sum_{t\in S_X}\! e_{X,t},\\
s_{\mathrm{all}}
  &= \sqrt{\frac{1}{N-1}\sum_{X\in\mathcal{X}} \ \sum_{t\in S_X}\! \Bigl(e_{X,t}-\bar e_{\mathrm{all}}\Bigr)^2 }.
\end{align*}
For percentage metrics let \(S_X^{(\%)}\) be as above and \(N^{(\%)}=\sum_{X\in\mathcal{X}} |S_X^{(\%)}|\). The pooled definition is
\[
\mathrm{MAPE}_{\mathrm{all}}
=\frac{100}{N^{(\%)}}\sum_{X\in\mathcal{X}}\ \sum_{t\in S_X^{(\%)}} \!\lvert r_{X,t}\rvert.
\]

We evaluate the module using a 60-day multi-asset price dataset. It covers several Solana assets with healthy trading volumes, identified by their SPL addresses, randomly sampled within the 60-day window and benchmarked against Birdeye. Each asset contributes about 40 to 100 observations. Each observation is the minute closest to the intended sampling time. We use this dataset only as a test bed to assess the robustness of the proposed pricing algorithm, with no fitting or parameter tuning. For each asset \(X\) and time \(t\), we query our pricing module and fetch the matching Birdeye value as ground truth to evaluate performance.

\begin{table}[H]
\renewcommand{\arraystretch}{1.3}
\setlength{\tabcolsep}{3.5pt} 
\caption{Benchmark accuracy against Birdeye (combined base): mean squared error (MSE), standard deviation (SD) of pointwise errors \(e_{X,t}=\widehat{P}_{X,t}-G_{X,t}\), and mean absolute percentage error (MAPE, \%).}
\label{tab:mse-sd-oneprice}
\centering
\scriptsize
\begin{tabular}{|c||c|c|c|c|}
\hline
\textbf{Asset \(X\)} & \(\boldsymbol{n_X}\) & \textbf{MSE} \(\big(\mathrm{MSE}_X\big)\) & \textbf{SD} \(\big(s_X\big)\) & \textbf{MAPE (\%)}\\
\hline
$X_1$  &  91 & \(1.75\times10^{-15}\) & 0.000 & 0.1 \\
$X_2$  & 102 & \(1.92\times10^{-6}\)  & 0.001 & 0.1 \\
$X_3$  &  94 & \(1.62\times10^{-6}\)  & 0.001 & 0.1 \\
$X_4$  &  97 & \(1.85\times10^{-3}\)  & 0.043 & 0.3 \\
$X_5$  &  86 & \(9.47\times10^{-5}\)  & 0.010 & 0.3 \\
$X_6$  &  90 & \(1.72\times10^{-4}\)  & 0.013 & 1.2 \\
$X_7$  &  91 & \(1.19\times10^{-8}\)  & 0.000 & 0.2 \\
$X_8$  &  89 & \(2.95\times10^{-1}\)  & 0.539 & 1.4 \\
$X_9$  &  94 & \(1.93\times10^{-7}\)  & 0.000 & 0.2 \\
$X_{10}$ & 36 & \(1.48\times10^{-6}\)  & 0.001 & 0.1 \\
\hline
\multirow{2}{*}{\textbf{Pooled (all)}} & \(\sum_X n_X\) & \(\mathrm{MSE}_{\mathrm{all}}\) & \(s_{\mathrm{all}}\) & \(\mathrm{MAPE}_{\mathrm{all}}\) \\
& 870 & \(3.04\times10^{-2}\) & 0.174 & 0.4 \\
\hline
\end{tabular}
\end{table}

This section shows our pricing algorithm matches Birdeye almost perfectly at one minute resolution. Across ten assets and \(N=870\) randomly sample pricing points, the pooled error is \(\mathrm{MAPE}_{\mathrm{all}}=0.4\%\), and the median per-asset MAPE is \(0.2\%\). Eight of ten assets are at or below \(0.3\%\) MAPE, and six are at or below \(0.2\%\). Several assets show essentially zero squared error (for example \(X_1\): \(1.75\times 10^{-15}\)), which indicates that, without any fitting or tuning, our slot based VWAP reproduces Birdeye’s ground truth to within rounding noise. Even on the highest-variance asset (\(X_8\)), accuracy remains strong with MAPE \(1.4\%\), while the others maintain very small variance. In short, the module delivers errors below one percent with default settings and achieves fidelity to the ground truth suitable for production use.

With the utility module complete, building the API integration and CI/CD pipeline is straightforward using deployment scripts (see Appendix~\ref{appendix:pricing-cicd}) . The code is available in the project’s GitHub repository.

\section{Discussions}
\label{sec:discussion}
The Solana pricing module provides a concrete, end to end demonstration that high quality on chain data can be reconstructed \emph{directly} from canonical blokchain data, without relying on premium third party endpoints, and at a fraction of the cost. In Section~\ref{case_study} we follow the complete LinkXplore flow (Fig.~\ref{fig:linkxplore-framework}), starting from identifying a market need, then moving through algorithm design and decoding utilities, implementation, cost and accuracy validation, and finally API integration. This walkthrough is not a merely an illustrative example but is backed by a production level engineering pipeline. In high-throughput settings, real time streaming introduces \emph{zero} marginal cost on our side, whereas comparable metered solutions grow proportionally with time (Fig.~\ref{fig:cost_rt}); for major events over relatively short historical windows, our total spend drops from hundreds or even thousands of dollars to only a few dollars (Fig.~\ref{fig:cost_event}). At the same time, accuracy remains at the level expected in industry: across 870 randomized evaluation points, the pooled error is $\mathrm{MAPE}_{\mathrm{all}}=0.4\%$ (Table~\ref{tab:mse-sd-oneprice}), which corresponds to at least ${\ge}99.5\%$ agreement in most cases.

While the added module significantly reduces costs under several popular scenarios, there are cases where its advantage narrows. If a user needs \emph{very sparse} points scattered across a \emph{long historical horizon}, the total cost of querying those points individually over RPC can approach roughly half of a commercial provider’s metered cost. In addition, large providers often optimize storage and thus achieve lower latency for cold history queries. This is, in effect, the classic “no free lunch”: optimizing on-chain utility performance often means taking on some extra engineering work of our own, for example building local indices, batching requests, adding simple caches, or allocating more storage. Our design leaves these choices to the deployer: keep the core open source and minimal, and people can add cost or latency optimizations only where the workload justifies them.

Apart from the outstanding performance of the pricing module, a natural question is whether our framework is tailored only to pricing. We emphasize that the framework is \emph{general}. The core submodule in Section~\ref{case_study}, the swap decoder, interacts with most of Solana’s low level information flows, including binary layouts, Token/Token22 transfers, program specific IDLs, router events, tracing of inner instructions, and sanity checks on signer direction. These are exactly the technical preliminaries required by many other data products. A concrete example is OHLCV candles, a feature that is commonly offered at a cost but can be implemented easily using the swapinfo utilities in the pricing module (see Appendix~\ref{appendixA}). Beyond OHLCV, the same adapter pattern supports several other endpoints, including (i) liquidity and pool state snapshots using the same decoders and base normalization, (ii) address-level P\&L and realized slippage, and (iii) risk and event monitors such as mint freezes and program upgrades. 

For academics and small- and medium-scale engineering teams, the immediate benefit is that many utilities previously gated by API bills can now be executed within a cheap RPC plan, with accuracy that tracks reputable references to within ${\sim}0.5\%$ in most tested cases. \emph{Cost-aware, reference-validated} framework modules encourage a healthy cycle in which community contributions converge into a stable API and remain affordable to the community. The framework’s development scheme converts “features” into \emph{well-tested modules} that others can reuse, inspect, and extend.

\section{Conclusion}
LinkXplore indicates that the supposed trade-off between cost and quality in blockchain data is much less substantial than commonly assumed. With careful design one can obtain high quality blockchain data at significantly lower cost. Since the development path is explicit and testable, the case study also serves as a practical template for constructing additional high-demand endpoints following the same approach.

Future work will focus on developing and implementing additional modules that cover a broader range of blockchain data types aligned with current industry demand, applying performance optimizations to existing modules to improve access speed and accuracy while further reducing cost, and maintaining these modules so they keep pace with changes across chains, such as program upgrades and new protocol deployments. Ultimately, we aim to make LinkXplore the leading open framework for serving high-quality, affordable blockchain data.


%

\appendices
\section{OHLCV Construction from Onchain Swaps}\label{appendixA}
This appendix specifies a concise, implementation ready procedure to build OHLCV candles for a target token \(T\) from decoded swap data.

\subsection{Per Trade Normalization (USD)}
For each swap at time \(t_i\) where \(T\) appears as \emph{TokenIn} or \emph{TokenOut} (not a routing hop), extract raw base unit amounts and decimals:
\[
A_T^{\text{raw}},\, d_T,\qquad A_C^{\text{raw}},\, d_C,
\]
and convert to UI units:
\[
a_T=\frac{A_T^{\text{raw}}}{10^{d_T}},\qquad a_C=\frac{A_C^{\text{raw}}}{10^{d_C}}.
\]
Define the per swap price in counter units per one target token:
\[
p_C=\frac{a_C}{a_T}.
\]
Convert to USD:
\[
p_i=
\begin{cases}
p_C, & \text{if } C \in \{\text{USDC, USDT}\},\\[4pt]
p_C \cdot s(t_i), & \text{if } C=\text{SOL,}\\
& \text{with } s(t_i)=\text{SOLUSD minute close},\\[4pt]
\text{discard}, & \text{otherwise}.
\end{cases}
\]
Define trade quantity and notional:
\[
q_i=a_T,\qquad u_i=p_i\,q_i.
\]
\textbf{Deduplication:} Use the transaction signature as key; if multiple valid \(T\!\leftrightarrow\!C\) legs exist in one signature, keep the leg with maximal \(u_i\).
The result is a time ordered trade stream \(\{(t_i,p_i,q_i,u_i)\}\) with \(p_i\) in USD per one token \(T\).

\subsection{Time Bucketing}
Choose interval \(\Delta\) (for example, 1 minute) and anchor
\[
\tau_0=\Big\lfloor \frac{t_{\text{start}}}{\Delta}\Big\rfloor \Delta.
\]
Assign trade \(i\) to bucket
\[
k=\Big\lfloor \frac{t_i-\tau_0}{\Delta}\Big\rfloor,\qquad \text{candle start } \tau_k=\tau_0+k\Delta.
\]

\subsection{Optional Robust Price Filter (Per Bucket)}
For bucket \(k\) with prices \(\{p_i\}\), let \(x_i=\ln p_i\) and \(m=\mathrm{median}(\{x_i\})\).
Keep trades satisfying the symmetric log fence with parameter \(r>1\) (for example, \(r=1.5\)):
\[
|x_i-m|\le \ln r
\;\;\Longleftrightarrow\;\;
r^{-1}\le \frac{p_i}{e^{m}}\le r.
\]
Apply the same mask to \((p_i,q_i,u_i)\).

\subsection{OHLCV Aggregation (Per Bucket)}
Let the remaining trades in bucket \(k\) be sorted by time \(t_{k,1}\le \cdots \le t_{k,n_k}\), with \((p_{k,j},q_{k,j},u_{k,j})\).
\[
\mathrm{Open}_k=
\begin{cases}
p_{k,1}, & n_k\ge 1,\\
\text{undefined}, & n_k=0.
\end{cases}
\]

\[
\mathrm{High}_k=\max_{1\le j\le n_k} p_{k,j}, \mathrm{Low}_k=\min_{1\le j\le n_k} p_{k,j}.
\]

\[
\mathrm{Close}_k=
\begin{cases}
p_{k,n_k}, & n_k\ge 1,\\
\text{undefined}, & n_k=0.
\end{cases}
\]

\[
\mathrm{VolToken}_k=\sum_{j=1}^{n_k} q_{k,j}, \mathrm{VolUSD}_k=\sum_{j=1}^{n_k} u_{k,j}.
\]

\textbf{Optional VWAP Close:}
\[
\mathrm{Close}^{\mathrm{VWAP}}_k=
\frac{\sum_{j=1}^{n_k} u_{k,j}}{\sum_{j=1}^{n_k} q_{k,j}}
\quad(\text{Open, High, and Low remain unchanged}).
\]

\subsection{Conventions and Edge Cases}
\begin{itemize}
  \item \textbf{Empty buckets} (\(n_k=0\)): set \(\mathrm{VolToken}_k=\mathrm{VolUSD}_k=0\); O, H, L, and C remain undefined (optionally carry the previous Close if continuity is required; do not invent volume).
  \item \textbf{Venue aggregation}: include any direct \(T\!\leftrightarrow\!C\) leg (not routing hops). This naturally aggregates across pools and routers.
  \item \textbf{SOL pricing cache}: cache \(s(t)\) by minute to avoid repeated SOLUSD lookups.
  \item \textbf{Precision}: normalize decimals before ratios; keep \(p_i,q_i,u_i\) in double precision or higher.
\end{itemize}

\subsection{Summary (One Liners)}
\[
p_i=
\begin{cases}
\frac{a_C}{a_T}, & \text{stable},\\[2pt]
\frac{a_C}{a_T}\cdot s(t_i), & \text{SOL}.
\end{cases}
\quad
q_i=a_T,\quad
u_i=p_i q_i,\quad
k=\Big\lfloor \frac{t_i-\tau_0}{\Delta}\Big\rfloor.
\]

\[
\mathrm{Open}_k=p_{k,1},\;
\mathrm{High}_k=\max p_{k,j},\;
\mathrm{Low}_k=\min p_{k,j},
\]

\[
\mathrm{Close}_k=p_{k,n_k},\;
\mathrm{VolToken}_k=\sum q_{k,j},\;
\mathrm{VolUSD}_k=\sum u_{k,j}.
\]

\[
|\ln p_i-\mathrm{median}(\ln p)|\le \ln r.
\]

\section{Pricing Module: API Integration CI/CD Pipeline}\label{appendix:pricing-cicd}
This appendix summarizes a pragmatic CI/CD pipeline used to build, test, ship, and promote the Pricing Module’s on-chain swap parsing API. The pipeline is script driven and portable across developer laptops and a small VM.

\subsection{Overview}
The flow is: \emph{local tests} $\rightarrow$ \emph{image build} $\rightarrow$ \emph{ship image tar to VM} $\rightarrow$ \emph{deploy (simple or blue--green)} $\rightarrow$ \emph{remote smoke}. Scripts are Bash, compose minimal dependencies, and expose tunables through \texttt{.env}.

\subsection{Local Test Harness (\texttt{run-tests.sh}, \texttt{deploy.sh})}
\begin{itemize}
  \item Builds a temporary Docker image \texttt{txparser:test-local} and runs a container bound to \texttt{127.0.0.1:8080}. Port conflicts are freed by stopping containers publishing 8080; non-Docker listeners abort with guidance.
  \item Health gate via \texttt{/healthz}. If the container is not healthy within a timeout, logs are printed and the run fails.
  \item Integration checks iterate \texttt{tests/cases.json}. Each case POSTs to \texttt{/parse} and passes when \texttt{swapInfo} is present and non-empty. Pacing is configurable.
  \item Optional holder sanity: GET \texttt{/holders?mint=...} for a mint list; this is informational and does not affect pass percentage.
  \item Summary threshold: deployment proceeds only if pass percentage $\ge$ \texttt{MIN\_PASS\_PCT} (default 50\%).
\end{itemize}

\subsection{Build and Ship Image Tar (\texttt{build-and-ship.sh})}
\begin{itemize}
  \item Builds a linux/amd64 image with \texttt{docker buildx}, tags it with a timestamp, then \texttt{docker save} to a tarball.
  \item Secure copy to the VM; prints only the tag for later reference.
  \item Requires \texttt{VM\_HOST}, \texttt{VM\_USER}. Optional names: \texttt{IMAGE\_NAME}, \texttt{REMOTE\_ARCHIVE}.
\end{itemize}

\subsection{Deployment Modes on VM}
\paragraph{Simple (\texttt{deploy-simple.sh}).}
\begin{itemize}
  \item \texttt{docker load} the tarball, discover the just-loaded tag, stop old container, and run the new one on \texttt{0.0.0.0:\$LIVE\_PORT} (default 8080).
  \item Optional \texttt{--env-file} (\texttt{\$HOME/txparser/.env}) injects RPC keys and service settings.
  \item Health checks on \texttt{/healthz} gate success.
\end{itemize}

\paragraph{Blue--Green (\texttt{deploy-bluegreen.sh}).}
\begin{itemize}
  \item Load the image, start \texttt{txparser\_new} on the staging port (\texttt{\$STAGE\_PORT}, default 8081) with restart policy and optional env file.
  \item Health check then smoke suite against the staging instance:
    \begin{enumerate}
      \item \emph{Swap smoke}: POST \texttt{/parse} for each signature in \texttt{~/txparser/tests/cases.json}. Each case may retry, with pacing.
      \item \emph{Holders smoke} (optional): GET \texttt{/holders?mint=\dots} for each mint in \texttt{holders.json}.
    \end{enumerate}
  \item If all pass, promote by renaming \texttt{txparser\_new} to \texttt{txparser}, rebind to the live port, and re-check \texttt{/healthz}. Failures cause immediate rollback.
\end{itemize}

\subsection{End-to-End Orchestration (\texttt{local-cicd.sh})}
\begin{enumerate}
  \item Run local tests (\texttt{run-tests.sh}).
  \item Build and ship the image tar (\texttt{build-and-ship.sh}) and capture the tag.
  \item Sync deploy scripts, test cases, optional holders, and the \texttt{.env} file to the VM.
  \item Preflight Docker on the VM (ensure daemon is active and enabled).
  \item Deploy via either simple or blue--green mode.
  \item Execute remote smoke (\texttt{smoke-remote.sh}) against the live port.
\end{enumerate}

\subsection{Remote Smoke Against Live (\texttt{smoke-remote.sh})}
\begin{itemize}
  \item Verifies \texttt{/healthz}.
  \item Replays swap cases on the live endpoint with bounded retries and pacing; marks overall failure if any case fails.
  \item Optionally checks holders for mints; failures contribute to overall status when enabled.
\end{itemize}

\subsection{Configuration and Safety}
\begin{itemize}
  \item All scripts accept environment variables from \texttt{.env} (RPC URLs, ports, pacing, thresholds). Sensitive keys are injected at runtime via \texttt{--env-file}; they are not baked into images.
  \item Health checks, explicit timeouts, and minimal retries prevent hangs. Every critical step has a clear failure path with logs.
  \item Blue--green mode isolates smoke validation from production traffic. Promotion occurs only after health and smoke succeed.
\end{itemize}

\subsection{Local Compose Helpers}
\begin{itemize}
  \item \texttt{up-local.sh} and \texttt{down-local.sh} provide a quick compose based spin up and teardown for development and manual exploration, with a short health wait.
\end{itemize}

\subsection{Operational Notes}
\begin{itemize}
  \item Ports: the local harness binds \texttt{127.0.0.1:8080}; the VM uses configurable live and staging ports; port 8080 is freed before tests.
  \item Observability: on health failures, recent container logs are surfaced automatically to speed up diagnosis.
  \item Idempotence: image discovery after \texttt{docker load} uses repository and tag matching; container names are consistent for promote and rollback actions.
\end{itemize}

\ifCLASSOPTIONcaptionsoff
  \newpage
\fi



%
\bibliographystyle{IEEEtran}              
\bibliography{bibtex/bib/IEEEexample} 

%




\end{document}